\documentstyle[prl,aps,epsf]{revtex}
\draft
\begin{document}
\twocolumn[\hsize\textwidth\columnwidth\hsize\csname
@twocolumnfalse\endcsname
\title{Electroweak Monopoles}
\author{Y. M. Cho}
\address{Asia Pacific Center for Theoretical Physics
\\and
\\ Department of Physics, Seoul National University, Seoul 151-742, Korea}
\author{Kyoungtae Kimm}
\address{Center for Theoretical Physics, 
 Seoul National University, Seoul 151-742, Korea}
\maketitle
\begin{abstract}
We present finite 
energy analytic monopole and dyon
solutions whose size is fixed by the electroweak scale.
Our result shows that genuine 
electroweak monopole and dyon could exist whose
mass scale is much smaller than the grand unification scale.
\end{abstract}
]

It has generally been believed that in the electroweak theory of
Weinberg and Salam there
exists no topological monopole of physical interest.
The basis for this ``non-existence theorem'' is, of course, that with
the spontaneous symmetry breaking the quotient space $SU(2) \times
U(1)/U(1)_{\rm em}$ allows no non-trivial second homotopy.
This belief, however, is unfounded. Indeed recently
Cho and Maison~\cite{Cho0} have established 
that the Weinberg-Salam model has exactly the same topological
structure as the Georgi-Glashow model,
and demonstrated the existence of a new type of 
monopole and dyon solutions in the standard Weinberg-Salam
model. 
This was based on the observation that 
the Weinberg-Salam model, with the hypercharge $U(1)$,
could be viewed as a gauged $CP^1$ model in which the (normalized)
Higgs doublet plays the role of the $CP^1$ field. So 
the Weinberg-Salam model does have exactly the same nontrivial
second homotopy as the Georgi-Glashow model which allows topological monopoles.
Originally the Cho-Maison solutions were obtained by a numerical integration, 
but now a mathematically
rigorous existence proof has been established which supports
the numerical results \cite{Yang}.

The Cho-Maison monopole may be viewed as a hybrid between the Dirac monopole
and the 't Hooft-Polyakov monopole, because it has a $U(1)$ point
singularity at the center even though the $SU(2)$ part is completely
regular. Consequently it carries an infinite energy so that 
at the classical level the mass of the monopole remains arbitrary. 
{\em A priori} there is nothing
wrong with this, but nevertheless one
may wonder whether one can have an analytic 
electroweak monopole which has a finite 
energy.
The purpose of this Letter is to show that this is indeed possible,
and to present analytic electroweak monopole and dyon solutions.
Clearly the new monopoles should have important physical
applications in the phenomenology of electroweak interaction.

Let us start with the Lagrangian of
the standard Weinberg-Salam model,
\begin{eqnarray}
{\cal L} &=& -|D_{\mu}\mbox{\boldmath $\phi$}|^2 
 -\frac{\lambda}{2}\Big(\mbox{\boldmath $\phi$}^\dagger\mbox{\boldmath $\phi$}
 -\frac{\mu^2}{\lambda}\Big)^2
 -\frac{1}{4}(\mbox{\boldmath $ F$}_{\mu\nu})^2 
 -\frac{1}{4}(G_{\mu\nu})^2 , \nonumber
\end{eqnarray}
\begin{eqnarray}
D_{\mu} \mbox{\boldmath $\mbox{\boldmath ${\phi}$}$}
&=& \Big(\partial_{\mu} 
  -i\frac{g}{2}\mbox{\boldmath $\tau$}\!\cdot\!\mbox{\boldmath $A$}_{\mu}
  -i\frac{g'}{2} B_{\mu}\Big) \mbox{\boldmath $\phi$}
   ,
\label{lag1}
\end{eqnarray}
where ${\bf \mbox{\boldmath $\phi$}}$ is the Higgs doublet, 
$\mbox{\boldmath $F$}_{\mu\nu}$
and
$G_{\mu\nu}$ 
are the gauge field strengths of $SU(2)$ and $U(1)$ with the
potentials $\mbox{\boldmath $A$}_{\mu}$ and $B_{\mu}$.
Now we choose the following static spherically symmetric ansatz
\begin{eqnarray}
\mbox{\boldmath$\phi$}&=&\frac{1}{\sqrt{2}}\rho(r)\xi(\theta,\varphi),
  \nonumber \\
\xi&=&i\left(\begin{array}{cc}
         \sin (\theta/2)\,\, e^{-i\varphi}\\
       - \cos(\theta/2)
      \end{array} \right), 
\hspace{5mm}{\mbox{\boldmath $\hat{\phi}$}} = \xi^{\dagger}
\mbox{\boldmath $\tau$} \xi = - \hat{r},
  \nonumber \\
\mbox{\boldmath $A$}_{\mu} 
&=& \frac{1}{g} A(r){\mbox{\boldmath $\hat{\phi}$}} \partial_{\mu}t
   +\frac{1}{g}(f(r)-1) {\mbox{\boldmath $\hat{\phi}$}} \times
    \partial_{\mu} {\mbox{\boldmath $\hat{\phi}$}} 
\label{ansatz1},\\
B_{\mu} &=&-\frac{1}{g'} B(r) \partial_{\mu}t -
            \frac{1}{g'}(1-\cos\theta) \partial_{\mu} \varphi, 
\nonumber
\end{eqnarray}
where $(t,r, \theta, \varphi)$ are the spherical coordinates.
Notice that the apparent string
singularity along the negative z-axis in $\xi$ and $B_{\mu}$ is a pure
gauge artifact which can easily be removed with a hypercharge $U(1)$
gauge transformation. Indeed 
one can easily exociate the string by making the hypercharge $U(1)$ 
bundle non-trivial. So the above ansatz describes a most
general spherically symmetric ansatz of a $SU(2) \times U(1)$ dyon.
Here we emphasize the importance of the non-trivial $U(1)$ 
degrees of freedom to
make the ansatz spherically symmetric. Without the extra $U(1)$ the
Higgs doublet does not allow a spherically symmetric ansatz.
This is because the spherical symmetry for 
the gauge field involves the embedding
of the radial isotropy group $SO(2)$ into the gauge group 
that requires the Higgs field to be
invariant under the $U(1)$ subgroup of $SU(2)$. This is possible
with a Higgs triplet, 
but not with a Higgs doublet \cite{Forg}. 

With the spherically symmetric ansatz (\ref{ansatz1}) 
and with the proper boundary conditions one can 
obtain  
the Cho-Maison dyon solution whose magnetic charge is 
given by $4\pi/e$~\cite{Cho0}.
The regular part of the solution 
looks very much like the Julia-Zee dyon solution~\cite{Julia},
except that it  has a non-trivial
$B-A$ which represents the neutral $Z$ boson content 
of the dyon solution. 
Of course the energy of the Cho-Maison solutions becomes infinite, 
due to the magnetic singularity at
the origin.
A simple way to make the energy finite is to introduce 
the gravitational interaction~\cite{Bais}.
But the gravitational interaction is not likely remove the singularity
at the origin.  

To construct the analytic
monopole and dyon solutions,
notice that a non-Abelian gauge theory in general
is nothing but a special type of an Abelian gauge theory 
which has a well-defined set of charged vector fields as its source.
This must be obvious, but this trivial observation reminds us the fact
that the finite energy non-Abelian monopoles are really nothing but the
Abelian monopoles whose singularity is regularized 
by the charged vector fields.
From this perspective one can try to 
make the energy of the above solutions
finite by introducing additional interactions and/or charged vector fields.
In the following we present two ways to achieve this goal.

\noindent{A)} {\bf Electromagnetic Regularization}

We could try to regularize
the magnetic singularity of the Cho-Maison solutions 
with a judicious choice of an extra 
electromagnetic interaction of the charged vector field
with the monopole. 
This regularization would provide 
a most economic way to make the energy of the Cho-Maison solution finite, 
because here we could use the already existing $W$ boson without
introducing a new source.
To show that this is indeed possible notice 
that in the unitary gauge the Lagrangian (\ref{lag1})
can be written as 
\begin{eqnarray}
{\cal L}
&=&
-\frac{1}{2}(\partial_\mu \rho)^2
-\frac{\lambda}{2}\Big(\frac{\rho^2}{2}-\frac{\mu^2}{\lambda}\Big)^2
-\frac{1}{4}(F_{\mu\nu})^2 -\frac{1}{4} (G_{\mu\nu})^2
\nonumber \\
&&
-\frac{1}{2}|D_\mu W_\nu -D_\nu W_\mu |^2 
+\frac{1}{4}g^2(W_\mu^* W_\nu -W_\nu^* W_\mu)^2 
\label{lag2}\\
&& 
+ig F_{\mu\nu} W_\mu^*W_\nu 
-\frac{1}{4}\rho^2\Big(g^2W_\mu^* W_\mu
 +\frac{1}{2}(g'B_\mu\!-\!gA_\mu)^2\Big),
\nonumber 
\end{eqnarray}
where
$W_\mu=\frac{1}{\sqrt2}(A^1_\mu+iA^2_\mu)$,
$A_\mu=A_\mu^3$, and
$D_\mu W_\nu =(\partial_\mu +ig A_\mu)W_\nu$. 
In this gauge the spherically symmetric ansatz
(\ref{ansatz1}) is written as  
\begin{eqnarray}
\rho &=&\rho(r),\nonumber \\
W_\mu&=& \frac{i}{g}\frac{f(r)}{\sqrt2}e^{i\varphi}
      (\partial_\mu \theta +i \sin\theta \partial_\mu \varphi),
\nonumber \\
A_\mu&=&-\frac{1}{g}A(r)\partial_\mu t 
          -\frac{1}{g}(1-\cos\theta)\partial_\mu \varphi,
\label{ansatz2}\\
B_\mu&=& -\frac{1}{g'} B(r)\partial_\mu t  
          -\frac{1}{g'} (1-\cos\theta) \partial_\mu\varphi.
\nonumber 
\end{eqnarray}
To regularize the Cho-Maison dyon, 
we now introduce an extra interaction
${\cal L}'$,
\begin{eqnarray}
{\cal L}'=i\alpha gF_{\mu\nu}W_\mu^* W_\nu 
               +\frac{\beta}{4}g^2(W_\mu^*W_\nu-W_\nu^*W_\mu)^2. 
\label{int1}
\end{eqnarray}
With this additional interaction the energy of the dyon is given by 
$E=E_0 +E_1$, 
where  
\begin{eqnarray}
E_0&=&
\frac{2\pi}{g^2}\int\limits_0^\infty
\frac{dr}{r^2}\bigg\{
\frac{g^2}{g'^2}+1-2(1+\alpha)f^2+(1+\beta)f^4
\bigg\}
   \nonumber,\\
E_1&=&\frac{4\pi}{g^2} \int\limits_0^\infty dr \bigg\{ 
\frac{g^2}{2}(r\dot\rho)^2
+\frac{g^2}{4} f^2\rho^2 +\frac{g^2r^2}{8} (B-A)^2 \rho^2 
\nonumber \\
&&
+\frac{\lambda g^2r^2}{2}\Big(\frac{\rho^2}{2}
-\frac{\mu^2}{\lambda}\Big)^2
+(\dot f)^2 
+\frac{1}{2}(r\dot A)^2 
\nonumber \\
&&
+\frac{g^2}{2g'^2}(r\dot B)^2 
+  f^2 A^2
\bigg\}.  
\end{eqnarray}
Clearly $E_1$ could be made  finite with a proper boundary condition, 
but notice that when $\alpha=\beta=0$, $E_0$ becomes infinite. 
To make $E_0$ finite we must require
\begin{eqnarray}
1+\frac{g^2}{g'^2}-2(1+\alpha) f^2(0)+(1+\beta) f^4(0)=0.
\end{eqnarray}
Furthermore to extremise the energy functional we must have 
\begin{eqnarray}
(1+\alpha)f(0)-(1+\beta) f^3(0) =0.
\end{eqnarray}
Thus we must have 
\begin{eqnarray}
\label{cond3}
&&\frac{(1+\alpha)^2}{1+\beta}
=1+\frac{g^2}{g'^2}=\frac{1}{\sin^2\theta_{\rm w}} ,
 \nonumber \\
&&f(0)=\frac{1}{\sqrt{(1+\alpha)\sin^2\theta_{\rm w}}},
\end{eqnarray}
where $\theta_{\rm w}$ is the Weinberg angle.
In general $f(0)$ can assume an arbitrary value depending on the choice of 
$\alpha$.
But notice that, except for $f(0)=1$,
the $SU(2)$ gauge field is not 
well-defined at the origin. This means that only when $f(0)=1$,
or equivalently only when $\alpha=\beta$, 
the solution becomes analytic everywhere
including the origin. So from now on we will assume $f(0)=1$. 
 
\begin{figure}
\epsfysize=5.5cm
\centerline{\epsffile{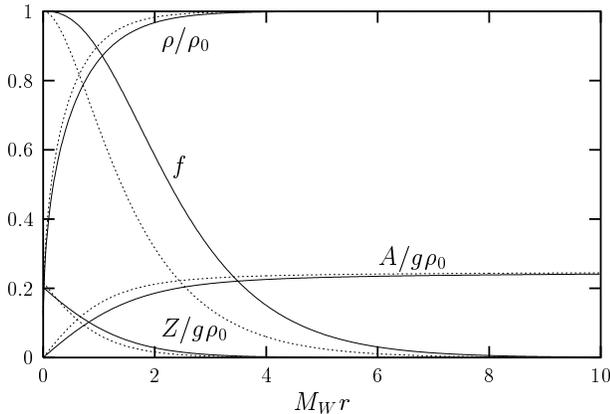}}
\caption{The electroweak dyon solutions. 
The solid line represents the finite energy dyon 
and dotted line represents the Cho-Maison dyon, where 
$Z=B-A$ and we have chosen $\sin^2\theta_{\rm w}=0.2325$,
$\lambda/g^2=0.5,$ and $A(\infty)=M_W/2$.}
\label{fig1}
\end{figure}

In this case the equations of motion that extremise the energy functional are
given by
\begin{eqnarray}
&&\ddot f - \frac{f^2 -1}{\sin^2\theta_{\rm w}r^2} f 
=\Big(\frac{g^2}{4}\rho^2-A^2 \Big) f ,
   \nonumber \\
&&\ddot \rho+\frac{2}{r}\dot\rho 
-\frac{f^2}{2r^2}\rho
=-\frac{1}{4}(B-A)^2 \rho 
+\lambda \Big(\frac{\rho^2}{2}
-\frac{\mu^2}{\lambda}\Big)\rho ,
 \nonumber \\
&&\ddot A +\frac{2}{r} \dot A 
- \frac{2f^2}{r^2} A 
=\frac{g^2}{4}(A-B)\rho^2 ,
\label{eqm3} \\
&&\ddot B+\frac{2}{r}\dot B 
=\frac{g'^2}{4}(B-A) \rho^2 ,
  \nonumber 
\end{eqnarray} 
which can be integrated 
with the boundary conditions 
\begin{eqnarray}
&&f(0)=1,~~A(0)=0 ,
~~B(0)=b_0,~~\rho(0)=0,\label{bound}\\
&&f(\infty)=0 ,~~A(\infty)=B(\infty),~~\rho(\infty)=\rho_0
                       =\sqrt{2\mu^2/\lambda}.
\nonumber
\end{eqnarray}
The result of the numerical integration for the finite energy dyon, 
together with the Cho-Maison dyon,
is shown in Fig.\ref{fig1}.
It is really remarkable that 
{\em the finite energy solutions look
almost identical to the Cho-Maison solutions,
even though they no longer have the singularity at the origin
and analytic everywhere}.

Clearly the energy of the above solutions 
must be of the order of the electroweak scale $M_W=g\rho_0/2$. 
Indeed for the monopole the energy with  
$\lambda/g^2=0.5$ is given by
\begin{eqnarray}
E=2.922\sin^2\theta_{\rm w}\times\frac{4\pi}{e^2}M_W.
\end{eqnarray}
This demonstrates that the finite energy solutions 
are really nothing but the regularized
Cho-Maison solutions which have a mass of the electroweak scale.

It is interesting to notice that for the monopole solution
we can find the Bogomol'nyi type energy bound 
if we add an extra term $\delta {\cal L}$ in the Lagrangian (\ref{lag2})
\begin{eqnarray}
\label{mass}
\delta{\cal L}=-\Big(\frac{1}{\sin^2\theta_{\rm w}}-\frac{1}{4}\Big)
                   g^2\rho^2W_\mu^* W_\mu .
\label{int2}
\end{eqnarray}
Notice that this amounts to changing the mass of the $W$ boson
from $g\rho_0/2$ to $g\rho_0/\sin\theta_{\rm w}$.
In this case we can show that with ansatz (\ref{ansatz2})
the energy of the monopole in the limit $\lambda=0$ is 
bounded from below by
\begin{eqnarray}
E\ge\frac{4\pi}{e}\rho(\infty) 
=\sin^2\theta_{\rm w}\frac{4\pi}{e^2}M_{W}.
\end{eqnarray}
{}Furthermore  this bound is saturated by the following 
Bogomol'nyi type equation, 
\begin{eqnarray}
&&\dot{f}+\frac{e}{\sin^2\theta_{\rm w}}\rho f=0,
\nonumber \\
&&\dot{\rho}-\frac{1}{er^2}(1-f^2) =0,
\label{self2}
\end{eqnarray}
which allows an analytic solution very much like the
Prasad-Sommerfield solution.
Notice that the energy of this solution has exactly the same form as the 
Prasad-Sommerfield monopole.
Obviously the solution is stable since it is the lowest energy configuration.

\noindent{B)} {\bf Embedding to $SU(2)\times SU(2)$}

 As we have noticed the origin of the infinite energy of the Cho-Maison
solutions was the magnetic singularity of $U(1)_{\rm em}$. On the other hand 
the ansatz (\ref{ansatz1}) also suggests that this singularity 
really originates from the magnetic part of the hypercharge
$U(1)$ field $B_\mu$.
So one could try to obtain a finite energy monopole solution  
by regularizing this hypercharge $U(1)$ singularity.
This could be done by enlarging the hypercharge $U(1)$
and embedding it to another $SU(2)$. 
This, of course, is same as introducing a hypercharged vector field
to the theory to regularize the $U(1)$ singularity.

To construct the desired solutions we  
introduce the hypercharge $SU(2)$ gauge field 
$\mbox{\boldmath$B$}_\mu$ and a scalar triplet $\bf\Phi$,
and consider the following 
Lagrangian 
\begin{eqnarray}
{\cal L} &=& 
-|D_{\mu}\mbox{\boldmath $\phi$}|^2 
-\frac{\lambda}{2}\Big( \mbox{\boldmath $\phi$}^\dagger\mbox{\boldmath $\phi$}
-\frac{\mu^2}{\lambda}\Big)^2
-\frac{1}{4} (\mbox{\boldmath $ F$}_{\mu\nu})^2 
\nonumber \\
&&
-\frac{1}{2}(\tilde{D}_\mu {\bf\Phi})^2
-\frac{\kappa}{4}\Big({\bf\Phi}^2-\frac{m^2}{\kappa}\Big)^2
-\frac{1}{4} (\mbox{\boldmath $ G$}_{\mu\nu})^2,
\label{lag3}
\end{eqnarray}
where $\tilde{D}_\mu {\bf\Phi}=
(\partial_\mu+g'\mbox{\boldmath$B$}_\mu\times)
{\bf\Phi}$. 
Now in the unitary gauge let
$X_\mu =(B_\mu^1+i B_\mu^2)/\sqrt{2}$,  
$B_\mu = B_\mu^3 $,
$\tilde{D}_\mu X_\nu=(\partial_\mu+ig' B_\mu)X_\nu$,
${\bf\Phi}= (0,0,\sigma)$,
and choose the static spherically symmetric ansatz
\begin{eqnarray}
\sigma &=&\sigma(r), \nonumber \\
X_\mu &=&\frac{i}{g'}\frac{h(r)}{\sqrt{2}}e^{i\varphi} (\partial_\mu \theta
+i\sin\theta\partial_\mu \varphi),
\label{ansatz3}
\end{eqnarray}
together with ({\ref{ansatz2}).
With this  
we obtain the following equations, 
\begin{eqnarray}
&&\ddot{f} - \frac{f^2-1}{r^2}f = 
              \Big(\frac{g^2}{4}\rho^2 - A^2\Big)f, 
  \nonumber\\
&&\ddot{\rho} + \frac{2}{r} \dot{\rho} - \frac{f^2}{2r^2}\rho
  =- \frac{1}{4}(B-A)^2\rho + \lambda\Big(\frac{\rho^2}{2} -
   \frac{\mu^2}{\lambda}\Big)\rho ,
  \nonumber \\
&&\ddot{A} + \frac{2}{r}\dot{A} -\frac{2f^2}{r^2}A = \frac{g^2}{4}
   \rho^2(A-B),
  \nonumber \\ 
&&\ddot h -\frac{h^2-1}{r^2} h =(g'^2\sigma^2-B^2) h ,
  \label{eom4}\\
&&\ddot\sigma +\frac{2}{r}\dot\sigma -\frac{2h^2}{r^2} \sigma
= \kappa\Big(\sigma^2-\frac{m^2}{\kappa}\Big)\sigma , 
 \nonumber \\
&&\ddot{B} + \frac{2}{r} \dot{B}- \frac{2h^2}{r^2} B
 =  \frac{g'^2}{4} \rho^2 (B-A). 
\nonumber  
\end{eqnarray}
Now with the boundary condition (\ref{bound}) and with
\begin{eqnarray}
&&h(0)=1,~~\sigma(0)=0,\nonumber \\
&&h(\infty)=0,~~\sigma(\infty)=\sigma_0=\sqrt{m^2/\kappa},
\end{eqnarray}
one may try to find the desired 
solution. 
Clearly the spontaneous symmetry breaking of the hypercharge $SU(2)$ at the 
infinity adds a new scale $M_X=g'\sigma_0$, 
an intermediate scale which lies somewhere between
the grand unification scale and the electroweak scale, to the theory. 
Now, consider the monopole solution and let $A=B=0$ for simplicity. 
Then in the limit $\lambda=\kappa=0$  
we obtain the solution shown in Fig.\ref{fig2} with $M_X=10M_W$,   
whose energy   is given by  
\begin{eqnarray}
E=(\cos^2\theta_{\rm w}+0.195\sin^2\theta_{\rm w})\frac{4\pi}{e^2}M_X.
\end{eqnarray}
Clearly the solution describes 
the Cho-Maison monopole whose singularity is regularized by
the hypercharge vector field $X_\mu$. 
\begin{figure}
\epsfysize=5.5cm
\centerline{\epsffile{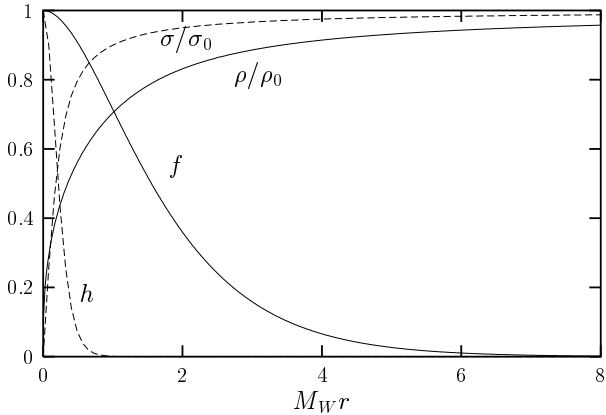}}
\caption{The $SU(2)\times SU(2)$ monopole solution.}
\label{fig2}
\end{figure}

It has generally been assumed that the finite
energy monopoles could exist only at the grand unification
scale~\cite{Dokos}. But our result tells that {\em there may exist 
a totally new class of electroweak
monopole and dyon whose mass is much smaller than the monopoles of the 
grand unification}.
Obviously the electroweak monopoles are topological solitons which 
must be stable. 

Strictly speaking the finite energy solutions are not the solutions of the
Weinberg-Salam model, because their existence requires a generalization of the 
model. But from the physical point of view there is no doubt that they should
be interpreted as the electroweak monopole and dyon, because 
they are really nothing but the regularized Cho-Maison solutions whose size is 
fixed at the electroweak scale. In spite of the fact that the Cho-Maison
solutions are obviously the solutions of the Weinberg-Salam model
one could try to object them as the electroweak dyons, 
under the presumption that the energy of the 
solutions could be made finite only at the grand unification scale.
Our work shows that 
this objection is groundless, 
and assures that 
it is not necessary for us to go to the 
grand unification scale to make the energy of the Cho-Maison solutions finite.

We close with the following remarks:

\noindent{1)} The electromagnetic regularization of the Dirac monopole 
with the charged vector fields is nothing new.
In fact it was this regularization which made the
energy of the 't Hooft-Polyakov monopole finite.
Furthermore it has been known that
the 't Hooft-Polyakov solution 
is the only analytic solution (with $\alpha=\beta=0$)
which one could obtain with this technique~\cite{klee}.
What we have shown in this paper is that the same
technique also works to regularize the Cho-Maison solutions, but
with nonvanishing $\alpha$ and $\beta$.

\noindent{2)} The introduction of the additional interactions (\ref{int1})
and (\ref{int2}) to the Lagrangian (\ref{lag1}) could spoil the 
renormalizability of the Weinberg-Salam model (although this issue has to 
be examined in more detail). How serious would this offense, however,
is not clear at this moment. Here we simply notice that
the introduction of a non-renormalizable interaction 
(like a gravitational interaction)  has been an acceptable practice
to study finite energy  classical solutions.

\noindent{3)} The embedding of the electroweak $SU(2)\times U(1)$ to
a larger $SU(2)\times SU(2)$ or 
$SU(2)\times SU(2)\times U(1)$ could 
naturally arise in the left-right symmetric grand unification
models, in particular in the $SO(10)$ grand unification, although
the embedding of the hypercharge $U(1)$ to a compact $SU(2)$
could turn out to be too simple to be realistic.
Independent of the details, however, our discussion strongly suggests  
that the electroweak monopoles at the intermediate scale
$M_X$ could be  possible in a realistic grand unification.

Certainly
the existence of the finite energy electroweak monopoles should have important
physical implications. 
Probably they could be the only finite energy topological monopoles
that one could ever hope to produce with the (future) accelerators.
A more detailed discussion of our work will be published in a separate
paper~\cite{cho97}.

\noindent{\bf Acknowledgments}

The work is supported in part by the Ministry of Education and by the Korean
Science and Engineering Foundation.


\begin{references}
\bibitem{Cho0} Y.M. Cho and D. Maison, Phys. Lett. {\bf B391}, 360 (1997).
\bibitem{Yang} Yisong Yang, Proc. Roy. Soc. {\bf A} (1997), in press.
\bibitem{Forg} P. Forg\'acs and N.S. Manton, Commun. Math. Phys.
           {\bf 72}, 15 (1980).

\bibitem{Julia} B. Julia and A. Zee, Phys. Rev. {\bf D11}, 2227 (1975);
             M.K. Prasad and C.M. Sommerfield, Phys. Rev. Lett.
         {\bf 35}, 760 (1975). 
\bibitem{Bais} F.A. Bais and R.J. Russell, Phys. Rev. {\bf D11}, 2692
                                (1975);
           Y.M. Cho and P.G.O. Freund, Phys. Rev. {\bf D12}, 1711 (1975);
           P. Breitenlohner, P.Forgacs, and D. Maison, Nucl. Phys. {\bf B383},
           357 (1992).
\bibitem{Dokos} C.P. Dokos and T.N. Tomaras, Phys. Rev. {\bf D21}, 2940
                            (1980);
           Y.M. Cho, Phys. Rev. Lett. {\bf 44}, 1115 (1980).
\bibitem{klee} K. Lee and E. Weinberg, Phys. Rev. Lett, {\bf 73}, 1203
                (1994);
               C. Lee and P. Yi, Phys. Lett. {\bf B348}, 100 (1995).
\bibitem{cho97} Y.M. Cho and K. Kimm, APCTP-97-12, to be published.
\end{references}
\end{document}